\documentclass[a4paper,fleqn]{cas-sc}
\usepackage[authoryear,longnamesfirst]{natbib}
\usepackage{setspace}
\usepackage{tabularx}
\usepackage{array}
\def\tsc#1{\csdef{#1}{\textsc{\lowercase{#1}}\xspace}}
\tsc{WGM}
\tsc{QE}
\tsc{EP}
\tsc{PMS}
\tsc{BEC}
\tsc{DE}

\begin{document}
\let\WriteBookmarks\relax
\def\floatpagepagefraction{1}
\def\textpagefraction{.001}
\shorttitle{Influence of asteroid shape elongation during a distant planetary encounter}
\shortauthors{Y Kim et~al.}

\title [mode = title]{The surface sensitivity of rubble-pile asteroids during a distant planetary encounter: Influence of asteroid shape elongation}        

\author[1]{Yaeji Kim}[type=editor,
                        auid=000,bioid=1,
                        orcid=0000-0002-9042-408X]
\cormark[1]

\ead{yzk0056@auburn.edu}

\address[1]{Department of Aerospace Engineering, Auburn University, 211 Davis Hall, Auburn, AL 36849-5338, USA}

\author[1]{Masatoshi Hirabayashi}

\author[2]{Richard P. Binzel}

\author[3]{Marina Brozović}

\author[4]{Daniel J. Scheeres}

\author[5]{Derek C. Richardson}

\address[2]{Department of Earth, Atmospheric, and Planetary Sciences, Massachusetts Institute of Technology, 77 Massachusetts Avenue, Cambridge, MA 02139 USA}

\address[3]{Jet Propulsion Laboratory, California Institute of Technology, 4800 Oak Grove Drive, Mail Stop 301-121, Pasadena, CA 91109-8099, USA}

\address[4]{Aerospace Engineering Department, University of Colorado Boulder, 429 UCB, Boulder, CO 80309, USA}

\address[5]{Department of Astronomy, University of Maryland, College Park, MD 20742, USA}

\nonumnote{* Corresponding author}

\begin{abstract}
We numerically investigate how an asteroid's elongation controls the sensitivity of its surface to tidal effects during a distant planetary encounter beyond the Roche limit. We analyze the surface slope and its variation by considering the shape elongation, as well as the spin period and orbital conditions. A more elongated asteroid tends to have a higher slope variation, while there may not be a monotonic increase of the total area having such a variation.
\end{abstract}

\begin{highlights}
\item We numerically investigate the influence of an asteroid’s elongation on the sensitivity of its surface to tidal effect during a distant planetary encounter beyond the Roche limit.
\item Objects spinning rapidly, but below the spin barrier, tend to experience a dramatic change of the surface slope during the distant encounters.
\item A more elongated asteroid tends to have a higher slope variation, while the asteroid elongation causes a non-linear increase in the total area having such a variation.
\end{highlights}
\maketitle
\doublespacing
\section{Introduction} 
S- and Q-type asteroids exhibit the compositional features of ordinary chondrites, while their spectral slope and absorption band are slightly different. S-type asteroids have steeper spectral slopes, weaker absorption bands near wavelengths between 1 $\mu$m and 2 $\mu$m, and lower albedos than Q-type asteroids \citep{binzel2010earth,chapman2004space,pieters2000space,vernazza2008compositional}. Space weathering driven by ion implantation and micrometeorite bombardments cause optical variations in material compositions \citep{chapman2004space, sasaki2001production}. For S- and Q-type asteroids, space weathering is considered to redden their surface materials. The timescale of space weathering may range between 10 ka and 1 Ma in the inner Solar System \citep{hapke2001space,strazzulla2005spectral}. For near-Earth asteroids (NEAs), the dynamical lifetime is 2 Ma \citep{bottke2002asteroids} and thus longer than the space weathering timescale. Without additional processes, Q-type asteroids should be altered to S-type asteroids and depleted continuously. However, this contradicts the high fraction of Q-type asteroids among NEAs ($\sim25\%$). One possible hypothesis is that redder S-type asteroids are resurfaced to become bluer Q-type asteroids \citep[e.g.,][]{binzel2010earth, nesvorny2010planetary}.

Possible resurfacing mechanisms include 1) planetary encounters with terrestrial planets \citep{binzel2010earth, carry2016spectral, demeo2014mars, marchi2006spectral, nesvorny2005evidence, nesvorny2010planetary}, 2) thermal fatigue by cyclic diurnal temperature variations \citep{delbo2014thermal,molaro2015grain,molaro2017thermally}, and 3) rotational instability driven by the Yarkovsky-O'Keefe-Radzievskii-Paddack (YORP) effect \citep{graves2018resurfacing}. Among these mechanisms, the tidal effect may be a critical driver that resurfaces rubble-pile asteroids. The distribution of Q-type asteroids correlates with the perihelion distance and the minimum orbit intersection distance, suggesting that encounters with massive planets may resurface asteroids enough to change their surface spectral properties \citep{binzel2010earth,marchi2006spectral, nesvorny2010planetary}. \cite{nesvorny2010planetary} proposed that Q-type asteroids may be resurfaced if the close encounter is within $\sim5$ planetary radii, thus outside the Roche limit ($\sim3.4$ planetary radii for an asteroid with a bulk density of 2.0 g cm$^{-3}$). These studies statistically showed that the resurfacing process of S-type asteroids is possibly related to their distant planetary encounters. However, the resurfacing mechanism is still not well understood.

Numerical studies have been reported for investigations of the tidal effects on rubble-pile asteroids during planetary encounters. They are in general divided in two categories; catastrophic, where an object is destroyed by the tides, and non-catastrophic, where either shape or surface or both are altered. For catastrophic disruptions, research has shown how small bodies that closely encounter planetary bodies are broken up due to strong tidal forces. The breakup conditions depend on small bodies' mechanical strength, material compositions, and orbital parameters \citep[e.g.,][]{dobrovolskis1990tidal, richardson1998tidal, sharma2006tidal, sridhar1992tidal}. In 1994, comet Shoemaker-Levy 9 encountered Jupiter and was broken into approximately 20 similar-sized fragments \citep{sekanina1994tidal}. \cite{scotti1993estimate} and \cite{Asphaug1994} modeled this event to better quantify catastrophic disruption processes during close tidal encounters. 

Tidal processes during distant encounters outside the Roche limit have also been studied in the literature. \cite{keane2014rejuvenating,Keane:lpsc2015} applied the theory of hill slope stability to evaluate the stability of asteroid regolith during the distant planetary flyby. They had two conclusions: rapidly rotating asteroids are highly likely to experience resurfacing, and the resurfacing process on asteroids may occur at larger flyby distance ($\sim$10 Earth Radii) than the previous estimation ($\sim$5 Earth Radii). In this paper, we extend their works to gain a more comprehensive understanding of resurfacing by adding an element that had not been discussed in the earlier works: an asteroid's elongation. We finally note that studies have shown limited changes in the shape and surface conditions of Apophis, which is going to closely flyby the Earth within 6 Earth radii \citep{demartini2019using,scheeres2005abrupt,yu2014numerical}. Our study will give further insight into the mechanism of mild tidal effects on a rubble-pile asteroid during a distant planetary encounter, which may be a critical source of Q-type asteroids.

In this study, we investigate the influence of a rubble-pile asteroid's elongation on surface sensitivity to tidal effects when it approaches the Earth outside the Roche limit by parameterizing the elongation with different planetary encounter conditions. We first demonstrate the resurfacing mechanism by considering the surface slope and its variation in Section \ref{Sec:resurfacing}. Our investigation and simulation settings are described in Sections \ref{Sec:Modeling} and \ref{Sec:setting}. In Section \ref{Sec:result}, the results of all conducted simulations are presented. We discuss our findings, potential issues of the current numerical model, and future work in Section \ref{Sec:Discussion}. 

\section{Resurfacing mechanism}\label{Sec:resurfacing}
Resurfacing processes remove a space-weathered, redder surface layer on S-type asteroids by exposing fresh materials beneath it. In situ observations by spacecraft revealed that the space weathered layer is likely very thin and correlated with the surface topography. Despite limited knowledge of space weathering on S-type asteroids, the sampled particles from NEA (25143) Itokawa suggested that the weathered thickness is only $\sim80$ nm \citep{noguchi2011incipient,noguchi2014space}, indicating that only a very-thin top surface layer had been affected by space weathering. We thus speculate that even mass movements at a tiny scale can induce color variations in top surfaces. Furthermore, \cite{sugita2019geomorphology, morota2020sample} reported on bluer, likely unweathered regions around the pole regions and the equatorial ridges of NEA (162171) Ryugu (C-type asteroid), although the space weathering mechanism may be different from that on S- and Q-type asteroids \citep{lantz2018space}.

We use two geophysical parameters to describe the surface sensitivity to possible granular flows: the surface slope and its variation. The surface slope describes how the surface element is tilted from the direction of its total acceleration combined with local gravity, the tidal effect, and the rotational effect on each surface element. The slope variation shows how the surface is affected by time-varying acceleration during the encounter. Using these parameters, we estimate the locations of where granular flows can occur. When the surface slope reaches its critical slope, i.e., the angle of repose, which is $35^\circ$ for a typical geological material without cohesion \citep{lambe2008soil}, granular flows may start to occur \citep{culling1960analytical,roering1999evidence}. Furthermore, if the slope variation is high enough to cause surface mobility, granular flows may occur for sub-critical slopes. \cite{ballouz2019surface} showed this mechanism operating on the surface of the Martian moon Phobos, leading to the resurfacing process; however, this mechanism may not be proper for the planetary encounter resurfacing because it is a non-periodic event. In our analysis, we use the slope variation to see how sensitive surfaces become during the distant planetary encounter. 

Mass movements in and on asteroids may be influenced by many different elements. Electrostatic forces may induce mass mobility \citep{hartzell2019dynamics}. Particle ejection may also occur by thermal fatigue \citep{molaro2019thermal}. Impact cratering can excavate fresh materials. Granular convection such as the Brazil-nut effect can transport fresh materials from the interior \citep{tancredi2012granular, yamada2014scaling}. Seismic wave propagation may enhance resurfacing \citep{richardson2005global,yasui2015experimental}. The consideration of these effects on mass movements is beyond our scope. Instead, we focus on surface flows driven by the tidal force, as well as the gravitational force and the centrifugal force to better understand how an asteroid's elongation affects the tidally induced resurfacing mechanism. 

\section{Modeling}
\label{Sec:Modeling}
Our model computes the surface slope ($\theta$) by using the following equation, 
\begin{equation}
\theta = \arccos \left ( \frac{- \vec{n} \cdot \vec{a_g}}{\| \vec{n} \| \|\vec{a_g} \|} \right ).
\end{equation}\label{slope}
where $\vec{a_g}$ is the net acceleration vector, and $\vec{n}$ is a normal vector to a surface element. The surface slope variation ($\delta\theta$) is the angle difference of a given element between the maximum surface slope during the flyby and the initial slope before the flyby. This quantity was used to analyze resurfacing on the martian moon, Phobos \citep{ballouz2019surface}.

Next, we discuss the net acceleration vector, $\vec{a_g}$, during a planetary encounter. Since we focus on the onset of the grain motion on an asteroid surface, particles are assumed to rest initially. The assumption yields $\dot {\vec{r}} = 0$ and thus excludes the Coriolis effect. This setting may provide a conservative condition for the occurrence of resurfacing. $\vec{a_g}$ is given as 
\begin{equation}
\vec{a}_g = -G\rho_A\int_{V_A}\frac{\Vec{r}}{r^3} dV_A - \frac{GM_E}{{R_c}^3}(\Vec{r}-\frac{3({\Vec{R}_c}+\Vec{r})\cdot\Vec{r}}{R_c^2}{\Vec{R}}_c) - \dot{\vec{{\omega}}}\times\vec{r} - \Vec{\omega}\times(\Vec{\omega}\times\Vec{r}). 
\label{Eq:acc}\end{equation}
In this equation, $G$ is the gravitational constant, $\rho_A$ is the asteroid's bulk density, $V_A$ is the asteroid's volume, $\vec{r}$ is the position of a surface element relative to the center of mass (COM) of the asteroid, $M_E$ is the mass of the planet, $\vec{R_c}$ is the position of the COM of the asteroid relative to that of the planet, $\Vec{\omega}$ is the asteroid's angular velocity vector, and $\dot{\Vec{\omega}}$ is the asteroid's angular acceleration vector. The first term is the acceleration by self-gravity, the second term describes the tidal acceleration, and the third and fourth terms represent the rotational effect. For self-gravity, we use an elliptical integral for computing the gravity field around a biaxial ellipsoid \citep{holsapple2001equilibrium}. 

For the rotational motion, we solve the following equation: 
\begin{equation}
[I]\Dot{\Vec{\omega}} + \Vec{\omega}\times[I]\Vec{\omega} = \frac{3GM_E}{R^5_c}\Vec{R_c}\times[I]\Vec{R_c}. \label{Eq:rot1}
\end{equation}
where $[I]$ is the moment of inertia matrix and computed by \cite{dobrovolskis1996inertia}. We also note the changes in the asteroid's orientation during the encounter. The transformation matrix, $[A]$, is given as
\begin{equation}
[\Dot{A}] = -[\Tilde{\omega}][A]. \label{Eq:rot2}
\end{equation}
The initial condition of [A] is fixed as an identity matrix. The evolution of [A] is eventually used to convert $\Vec{R_c}$, which is a vector in the planet-centered inertia reference frame, to the one in the body-fixed frame in Equation (\ref{Eq:acc}). Here, $[\tilde \omega]$ is the skew matrix of $\vec{\omega}$, which is given as
\begin{equation}
\Tilde{\omega} = \begin{bmatrix}
0 & -\omega_3 & \omega_2 \\
\omega_3 & 0 & -\omega_1 \\
-\omega_2 & \omega_1 & 0
\end{bmatrix}
\label{Eq:rot3}
\end{equation}

We use a Runge-Kutta 4th order integrator with a constant time-step to propagate $\vec{\omega}$ and $[A]$ over a considered time frame. $\vec{R_c}$ is described in the body-fixed frame centered at the COM of the planetary body, where the axes are aligned with the asteroid's moment of inertia axes. To determine $\vec{R_c}$, we design a hyperbolic trajectory with a given periapsis and a constant encounter velocity and obtain a list of $\Vec{R_c}$ within a considered time span with a proper time step (see Section \ref{Sec:setting}). This list is incorporated into the integrator that solves Equations (\ref{Eq:rot1}) and (\ref{Eq:rot2}) to describe the rotational motion. At last, all computed parameters are applied to Equation (\ref{Eq:acc}) to obtain $\vec{a}_g$ of each element on the asteroid. We developed a MATLAB program package for this simulation process. 

\section{Simulation settings}
\label{Sec:setting}
\begin{table}[width=\textwidth,cols=2,pos=b]
\centering
\caption{Physical parameters and shape model information used in our simulations.}\label{T1}
\begin{tabular*}{.8\textwidth}{@{} CCCC@{} }
\toprule
Parameter & Values \\
\midrule
Earth-like planet mass ($M_E$) & 5.97 $\times$ $10^{24}$ $kg$ \\
Asteroid's volume ($V_A$) & 1.98 $\times$ $10^{7}$ $m^3$ \\
Asteroid's bulk density ($\rho_A$) & 2.0 $g$ $cm^{-3}$ \\
Simulation length & < 20 $Earth Radii$\\
Simulation time step & 5 $sec$ \\
Encounter speed & 15 $km$ $s^{-1}$\\
Angle of repose of top surface layers & 35$^\circ{}$ \\
Number of facets &  10440\\
Mean facet size & 35 $m^{3}$ ($\sim$0.01$\%$ of the entire surface)\\ 
\bottomrule
\end{tabular*}
\end{table}
We focus on the slope variation on uniformly rotating biaxial ellipsoids with different aspect ratios ($AR$). Here, $AR$ is defined as a ratio of the major axis to the minor axis. The volume of the shape models is set identical to that of a sphere having an equivalent diameter of 340 m, which is a typical size of rubble pile asteroids. A bulk density is assumed to be a constant value of 2.0 g cm$^{-3}$, which is consistent with that of Itokawa (S-type)  \citep{fujiwara2006rubble}. In each hyperbolic trajectory, the initial and final points of a test asteroid are 20 Earth-like planet radii ($\sim$6371 km for one Earth-like planet radius) away from the planet center, and thus the tidal effect is negligible at those points. The encounter speed is set to 15 km s$^{-1}$ for all the cases \citep{bottke1994collisional}, while we consider three hyperbolic trajectories with different periapses in the range of plausible resurfacing distances (3.5, 5, and 10 planet radii) \citep{keane2014rejuvenating}. The total length of simulation is approximately 4.5 hours with a time step of 5 seconds. In all test cases, the body is set to rotate along their maximum moment of inertia axes, and its spin axis is perpendicular to the orbital plane. The angle of the repose is set to be 35$^\circ$.

\section{Results} \label{Sec:result}
We introduce the nomenclature used in this session. To characterize the area affected by high slope variation ($\delta \theta$), we introduce two characteristic regions: Regions I and II. Region I is an area whose surface slope is higher than the angle of repose of $35^\circ$ before the encounter. Region II, on the other hand, is an area that has the slope exceeding the angle of repose during the encounter. We also use the area difference between Regions I and II to indicate the surface regions that reach above the angle of repose due to the tidal effect. A total of 33 simulations are performed to analyze how $\delta \theta$ is affected by the asteroid elongation during distant encounters.  

\begin{figure}
	\centering
		\includegraphics[width=\textwidth]{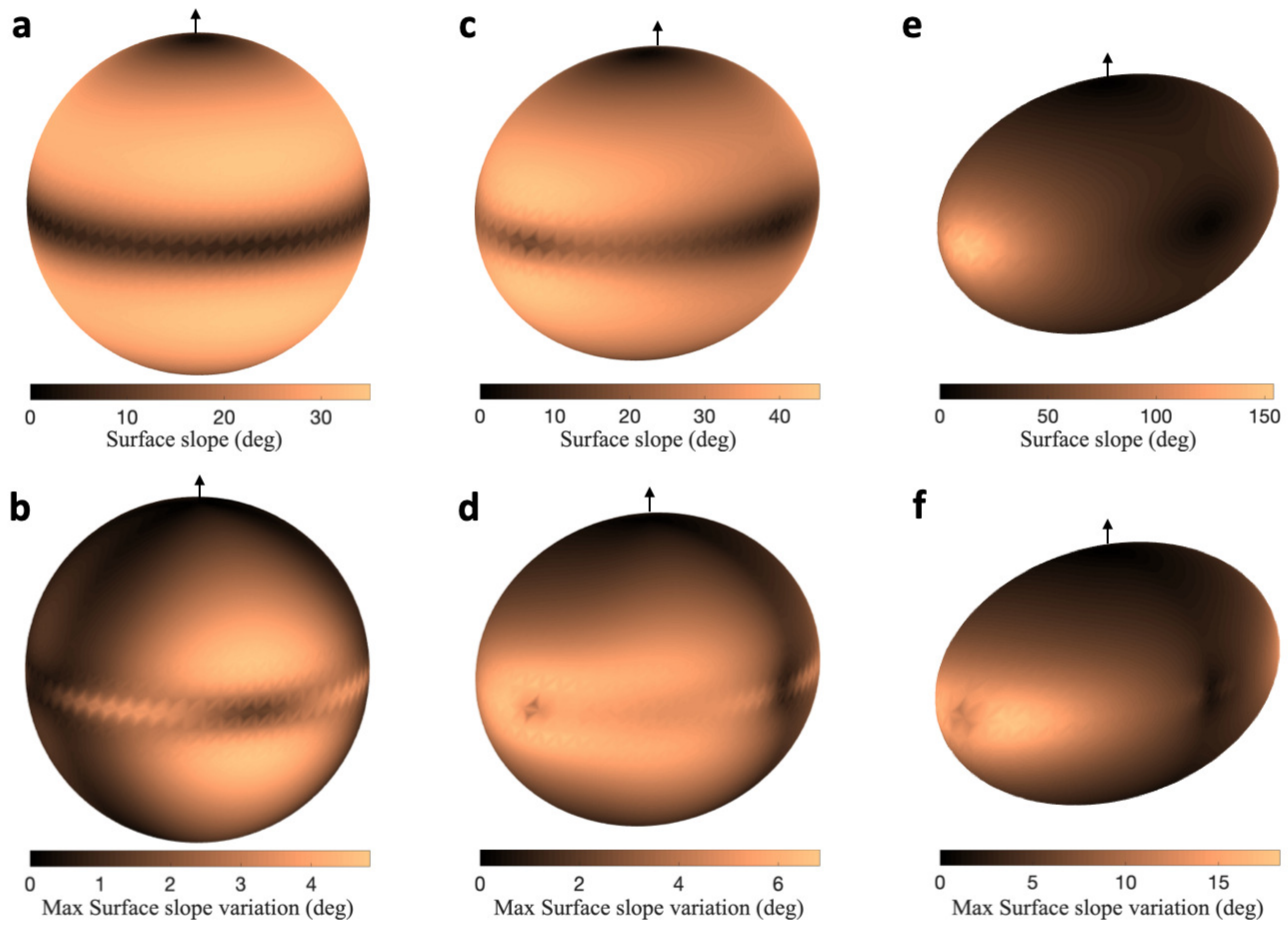}
	\caption{Maximum surface slope and slope variations for each shape. The top panels (a, c, and e) show the surface slopes, which are affected by the tidal effect, during the planetary encounter with the periapsis of 5 Earth radii and the spin period of 2.8 hr. The bottom panels (b, d, and f) illustrate the maximum surface slope variation during this encounter. The aspect ratio of each column is 1, 0.8, and 0.57, respectively. All plots are projected from a longitude of $130^\circ$ and a latitude of $15^\circ$. The black arrows denote the spin axes.}
	\label{FIG:1}
\end{figure}

We consider three cases of the spin periods, 1.5, 2.8, and 3.1 hr. These spin periods are selected to demonstrate high $\delta \theta$ variations around the spin barrier, where gravitational aggregates may fall apart if they do not have tensile strength \citep{pravec2006nea}. Given the spin periods defined, the test body may have high slopes even before a tidal encounter (i.e., Region I) in some cases. The 1.5 hr spin period case causes all the $AR$ cases to have Region I across the entire surface. Highly elongated shapes with lower $AR$s have Region I at even longer spin periods. At a spin period of 2.8 hr (see Figure \ref{FIG:1}), shapes with $AR=0.8$ and $AR=0.57$ have Region I at middle latitudes around the longest axis. At a spin period of 3.1 hr, the $AR=0.57$ shape still has Region I, which would be sensitive to granular flow. We emphasize that the reason for this selection is to illustrate the mechanisms of the slope variation due to the elongation and show its transition around the spin barrier. Thus, these conditions allow for visualizing how shape elongation controls surface slope variations, enhanced by rotation, during a distant encounter. 

To show how the elongation affects $\delta \theta$, Figure \ref{FIG:1} illustrates sample cases in which asteroids with $AR$s of 1.0, 0.8, and 0.57 reach 5 Earth radii with a spin period of 2.8 hr. The results show that $\delta \theta$ becomes higher as $AR$ is lower. When the asteroid has $AR=1$, high $\delta\theta$ occurs at a latitude of $\sim$18 deg. The high $\delta\theta$ regions always face the planet because of its axisymmetric shape. When $AR=0.8$, high $\delta \theta$ regions tend to be distributed widely around the edges along the longest axis. When $AR=0.57$, higher $\delta \theta$ concentrates on the edges.  

\begin{table}[width=\textwidth,cols=7,pos=ht]
\centering
\caption{Surface slope data. $\delta\theta$ shows the minimum and maximum value at Region II, excluding Region I. If this value is unchanged (i.e., Region I = Region II), $\delta\theta$ shows the slope variation across the entire surface with a parenthesis. The increase rate defines the area difference between Region II and Region I. For the 2.8 hr case, five $AR$ cases are introduced to illustrate the variations in the increase rate.}\label{T2}
\begin{tabular*}{.9\textwidth}{CCCCCCC}
\toprule
\multicolumn{2}{C}{Simulation Setting} & \multicolumn{2}{C}{Pre-encounter Time}&
\multicolumn{3}{C}{Planetary Encounter Time} \\ \cline{1-2} \cline{3-4} \cline{5-7}
\multicolumn{1}{C}{$AR$}&\multicolumn{1}{C}{Periapsis ($R_{E}$)}&\multicolumn{1}{C}{Mean slope ($^\circ$)}&\multicolumn{1}{C}{Region I ($\%$)}&\multicolumn{1}{C}{Region II
($\%$)}&\multicolumn{1}{C}{Increase Rate
($\%$)}&\multicolumn{1}{C}{$\delta\theta$ ($^\circ$)}\\
\midrule
\multicolumn{2}{L}{\textit{Spin period - 1.5 hr}}\\
1.0 & 3.5 & 83.2 & 97 & 97.4 & 0.4 & 0.47 - 1.94\\
0.8 & 3.5 & 82.7 & 96.2 & 96.2 & 0 & (0.20 - 4.68)\\
0.57 & 3.5 & 82.3 & 96.6 & 96.6 & 0 &2.50 - 3.17\\
1.0 & 5 & 83.2 & 97 & 97.4 & 0.4 & 0.31 - 0.69 \\
0.8 & 5 & 82.7 & 96.2 & 96.2 & 0 & (0.06 - 1.62) \\
0.57 & 5 & 82.3 & 96.6 & 96.6 & 0 & 0.63 - 0.80 \\
1.0 & 10 & 83.2 & 97 & 97.4 & 0.4 & 0.06 - 0.09 \\
0.8 & 10 & 82.7 & 96.2 & 96.2 & 0 & (0.01 - 0.23) \\
0.57 & 10 & 82.3 & 96.6 & 96.6 & 0 & (0.01 - 0.30) \\
\multicolumn{2}{L}{\textit{Spin period - 2.8 hr}}\\
1.0 & 3.5 & 22.3 & 0 & 22.7 & 22.7 & 4.12 - 16.23\\
0.9 & 3.5 & 22.7 & 0 & 31.5 & 31.5 & 3.13 - 19.89\\
0.8 & 3.5 & 24 & 10.2 & 46.9 & 36.7 & 5.12 - 34.48\\
0.7 & 3.5 & 26.7 & 22.1 & 55 & 32.9 & 6.89 - 148.83\\
0.57 & 3.5 & 33.1 & 33.7 & 63.4 & 29.7 & 9.14 - 38.41\\
1.0 & 5 & 22.3 & 0 & 0.7 & 0.7 & 3.96 - 4.22\\
0.9 & 5 & 22.7 & 0 & 6.9 & 6.9 & 1.62 - 4.67\\
0.8 & 5 & 24 & 10.2 & 20.9 & 10.7 & 2.21 - 6.92 \\
0.7 & 5 & 26.7 & 22.1 & 33.2 & 11.1 & 2.88 - 34.24\\
0.57 & 5 & 33.1 & 33.7 & 43.6 & 9.9 & 3.80 - 11.22 \\
1.0 & 10 & 22.3 & 0 & 0 & 0 & (0.01 - 0.53)\\
0.9 & 10 & 22.7 & 0 & 0 & 0 & (0.01 - 0.74)\\
0.8 & 10 & 24 & 10.2 & 11.8 & 1.6 & 0.31 - 0.75 \\
0.7 & 10 & 26.7 & 22.1 & 23.5 & 1.4 & 0.34 - 2.66 \\
0.57 & 10 & 33.1 & 33.7 & 34.8 & 1.1 & 0.39 - 0.98 \\
\multicolumn{2}{L}{\textit{Spin period - 3.1 hr}}\\
1.0 & 3.5 & 16.6 & 0 & 0 & 0 &  (0.04 - 13.08)\\
0.8 & 3.5 & 16.9 & 0 & 0 & 0 &  (0.10 - 11.08)\\
0.57 & 3.5 & 22.5 & 15.4 & 29.7 & 14.3 & 7.13 - 21.88\\
1.0 & 5 & 16.6 & 0 & 0 & 0 &  (0.02 - 3.72)\\
0.8 & 5 & 16.9 & 0 & 0 & 0 &  (0.02 - 4.73)\\
0.57 & 5 & 22.5 & 15.4 & 17.7 & 2.3 & 2.41 - 8.36\\
1.0 & 10 & 16.6 & 0 & 0 & 0 &  (0.01 - 0.39)\\
0.8 & 10 & 16.9 & 0 & 0 & 0 &  (0.01 - 0.60)\\
0.57 & 10 & 22.5 & 15.4 & 16.8 & 1.4 & 0.78 - 1.72\\
\bottomrule
\end{tabular*}
\end{table}

Table \ref{T2} shows the results of the cases simulations. We first discuss the cases of 1.5 hr and 3.1 hr. When the spin period is 1.5 hr, since the centrifugal force is already dominant, Region I spreads over the almost entire regions for any shapes. This extreme condition does not cause the slope to change significantly. When the spin period is 3.1 hr, self-gravity starts to be dominant, and tidal acceleration does not influence the slope variation remarkably. The $AR=0.57$ case still shows non-zero area difference between Region II and Region I at the asteroid's edge along the longest axis, where self-gravitational acceleration and rotational acceleration are comparable.

When the spin period is 2.8 hr, rotational acceleration becomes comparable to self-gravitational acceleration while some locations on elongated shapes may still be dominantly influenced by centrifugal forces. We show more simulation cases for this spin period case in Table \ref{T2}, compared to other spin periods, to examine the slope variation depending on the elongation. Region II becomes wider than Region I, strongly depending on the elongation. With the decrease of $AR$, the area difference between Region II and Region I (see the Increase Rate in Table \ref{T2}) increases in the range of $AR$ from 1 to $\sim$0.8 but decreases from $\sim$0.8 to 0.57. Consider when the periapsis is 3.5 Earth radii. For $AR = 1$, Region II is $\sim$23\% broader than Region I. The area difference becomes $\sim$37\% for $AR=0.8$ and then gradually decreases to $\sim$30\% for $AR=0.57$, which is still higher than the spherical case. This similar trend is observed when the periapsis is 5 and 10 planet radii. The decrease of the area difference from $AR=0.8$ to $AR=0.57$ results from the fact that a more elongated shape has a limited region in which a high slope variation occurs because of its narrower edge. As the periapsis becomes distant, tidal acceleration becomes smaller, leading to the decrease of the slope variation. 

Our results are consistent with \cite{keane2014rejuvenating, Keane:lpsc2015,yu2014numerical} in terms of the influence of the spin period and orbit on $\delta \theta$. Importantly, we newly address how the elongation impacts the slope variation during distant encounters. If the shape is moderately elongated, broader areas may have reasonable slope variations, which may cause the surface slope to reach the angle of repose. If the shape is highly elongated, the slope variation becomes higher although the affected area may be limited. While our parameter analysis is still coarse, the 2.8 hr case implies that $AR=0.8$ has the most noticeable change between Regions II and I because the total acceleration of rotation and self-gravity becomes small across the broader regions, the tidal acceleration can easily change their slope conditions.

\section{Discussion}\label{Sec:Discussion}
We numerically showed how the shape elongation affects the slope variations and the location of sensitive regions during distant encounters. We observed two critical features. First, the shape elongation enhances the slope variation. High slope variations are observed when the shape is elongated. When an asteroid spins at a spin period at which centrifugal and gravity acceleration are comparable, tidal acceleration can dramatically change the surface slope. Thus, an elongated body tends to be more exposed to such a dramatic change. Second, the increase of the area whose surface slope exceeds 35$^\circ$ during a distant encounter (Region II - Region I) depends on shape elongation. As the shape is more elongated from the spherical condition, the area that reaches the angle of repose of 35$^\circ$ increases. However, after the shape reaches a certain elongation (in our simulation, it was $\sim$0.8), the area difference, Region II - Region I, starts decreasing due to the limited area of high slope variations. This trend is reasonable because the slope condition on a highly elongated body is clearly separated into gravitational and centrifugal acceleration-dominant regions along the longest axis, causing "fewer" areas to have high slope variations. From these two trends, we conclude that elongation is a strong contributor to the surface slope variation during a distant encounter.

The shape elongation of NEAs has been reported by observational investigations. The statistical characterizations of light-curve amplitudes by \cite{mcneill2019constraining} showed that the shape elongation of NEAs may be between 0.6 and 0.8 if the objects are assumed to be a population of prolate spheroids. Furthermore, there is an apparent excess of fast rotators in the NEAs with D > 200 m near the spin barrier. Our study implies that if the fast spinning asteroids experience distant encounters, high slope variations may occur, and their elongation may control the affected area. This interpretation is consistent with the arguments of earlier works \citep[e.g.,][]{Keane:lpsc2015}, and we emphasize that our study newly analyzes the influence of the elongation on slope variations. 

We point out potential issues with our current numerical model. First, this model does not take into account cohesion, which may prevent resurfacing at fast spin but enhance its magnitude if resurfacing occurs. However, the surface slope without cohesion is still a meaningful parameter based on recent observations \citep[e.g.,][]{sugita2019geomorphology,ballouz2019surface,fujiwara2006rubble,willner2014phobos} that thin top-surface layers of an asteroid may be covered with weak, cohesionless materials. Second, the shape of an asteroid is assumed to be rigid in our model; in other words, its deformation is ignored. If deformation occurs, the location of resurfaced regions and the magnitude of resurfacing are likely to change due to the variation of $[I]$ and $\vec{r}$. Third, in the performed simulations, we did not parameterize the asteroid's spin orientation. Depending the direction of the spin axis, the tidal effect may affect the rotational motion differently \citep{scheeres2001changes}, causing variations in surface acceleration. Thus, the slope distribution may evolve differently due to the rotational conditions during distant encounters. Fourth, we only considered biaxial ellipsoids for our asteroid shapes, which only represent an approximate subset of true shapes of NEAs. Lastly, this study only estimated incipient conditions when surface regolith could be set in motion but did not give the magnitude of the surface mobility. We leave these issues as critical elements to be solved in future work.

\section*{Acknowledgement}
Y.K. and M.H. acknowledge support from Auburn University's Intramural Grant Program.
\bibliographystyle{apalike}

\end{document}